\documentclass[floatfix,showpacs,preprintnumbers,amsmath,amssymb,aps,twocolumn,superscriptaddress,prl,10pt]{revtex4-1}
\usepackage{graphicx}
\usepackage{color}
\usepackage{multirow}
\newcommand{\cmplxi}{\text{i}}
\newcommand{\neweqnline}{\nonumber\\}
\newcommand{\punc}[1]{\,#1}
\newcommand{\vecgrk}[1]{\boldsymbol{#1}}
\newcommand{\tr}{\operatorname{Tr}}
\newcommand{\e}[1]{\text{e}^{#1}}

\newcommand{\figref}[1]{Fig.~\ref{#1}}
\newcommand{\eqnref}[1]{Eqn.~\eqref{#1}}
\newcommand{\tabref}[1]{Table~\eqref{#1}}

\renewcommand{\vec}[1]{{\bf#1}}

\begin{document}

\title{Itinerant ferromagnetism with finite ranged interactions}
\author{C.W.~von~Keyserlingk}
\affiliation{Rudolf Peierls Centre for Theoretical Physics, 1 Keble Road, Oxford, OX1~3NP, United Kingdom}
\author{G.J.~Conduit}
\email{gjc29@cam.ac.uk}
\affiliation{Cavendish Laboratory, J.J.~Thomson Avenue, Cambridge, CB3~0HE, United Kingdom}

\date{\today}

\begin{abstract}
  Quantum fluctuations are of central importance in itinerant ferromagnets;
  in the case of the Stoner Hamiltonian, with contact interactions, they
  deliver a rich phase diagram featuring a first order ferromagnetic
  transition preempted by a spin spiral and a paired density wave. However,
  to date all analyses of fluctuation corrections neglect the finite ranged
  nature of the Coulomb interaction. We develop the formalism to consider
  the effects of fluctuations with a realistic screened Coulomb potential.
  The finite ranged interaction suppresses the tricritical point temperature
  of the first order ferromagnetic transition, bringing theory into line
  with experiment, whilst retaining the exotic spin spiral and paired
  density wave. In an ultracold atomic gas a finite ranged interaction damps
  the competing molecular instability, permitting the observation of
  ferromagnetic correlations.
\end{abstract}

\maketitle

Metallic systems in the vicinity of second order phase transitions display
remarkable quantum critical phenomena~\cite{Hertz}. However, in several
metals~\cite{Uhlarz04,Huxley00,Rojas11,Pfleiderer01,OteroLeal08} quantum
criticality gives way to a first order transition, spatially modulated
magnetic order~\cite{Lausberg12,Borzi07,Wu11}, and a p-wave superconducting
instability~\cite{Saxena00,Huxley01,Watanabe02,Aoki01,Huy07}.  Starting from
a minimal Hamiltonian of freely dispersing electrons with a repulsive
contact interaction, theory indicates that soft magnetic fluctuations
transverse to the magnetic order drive the first order
transition~\cite{Belitz:1997cr,Efremov:2008oq,Maslov:2009kl,Rech:2006nx,Conduit08,Conduit09,Kirkpatrick:2012bs},
spin spiral phase~\cite{Conduit09}, and the p-wave superconducting
instability~\cite{Conduit13}. However, there is a significant discrepancy:
experiments typically show a tricritical temperature of
$T_{\text{c}}\approx0.02T_{\text{F}}$~\cite{Hooper04,Borzi07,Wu11} (in
Sr$_3$Ru$_2$O$_7$), while theory predicts $T_{\text{c}}\approx0.3
T_{\text{F}}$~\cite{Conduit09}.  Here we demonstrate both analytically and
numerically that this discrepancy can be resolved by using the realistic
screened Coulomb potential rather than the ubiquitous contact potential.

A cold atom gas presents an alternative forum to explore the Stoner
Hamiltonian. Attempts to observe itinerant ferromagnetism in a cold atom gas
have been
thwarted~\cite{Jo09,Duine05,Conduit10,Conduit10ii,Conduit10iii,Massignan11,vonKeyserlingk11}
by a competing instability to a molecular bound
state~\cite{Pekker10}. Motivated by the suppression of pairing in a narrow
Feshbach resonance~\cite{Pekker11,Kohstall12}, here we demonstrate how a
finite interaction range removes pairing in a region of stable ferromagnetic
order.

In this paper we study itinerant ferromagnetism in the presence of a finite
ranged repulsive interaction. We extend an analytical fluctuation correction
formalism that has already delivered a phase diagram containing a first
order ferromagnetic phase transition~\cite{Conduit08}, spiral
phase~\cite{Conduit09}, and p-wave superconducting
instability~\cite{Conduit13} to now include a finite ranged interaction and
calculate the phase diagram. Second, we perform complementary Quantum Monte
Carlo calculations to verify the zero temperature behavior. Finite ranged
interactions suppress the tricritical point temperature. Finally, we focus
on the cold atom gas with finite ranged interactions that displays
ferromagnetic order whilst circumventing the competing pairing process.

\section{Formalism}

To study itinerant ferromagnetism in the presence of a finite ranged
interaction we focus on the idealized Hamiltonian
\begin{align}
 &\hat{H}=\!\!\!\!\!\sum_{\vec{p},\sigma\in\{\uparrow,\downarrow\}}\!\!\!
 \xi_{\vec{p} \sigma}c^{\dagger}_{\vec{p},\sigma}c_{\vec{p},\sigma}\neweqnline
 &-\!\!\!\sum_{\vec{p}_{\uparrow},\vec{p}_{\downarrow},\vec{q}}\!\!\!
 g(\vec{p}_{\uparrow}\!-\!\vec{p}_{\downarrow},\vec{q})
 c^{\dagger}_{p_{\uparrow}\!-\!q\!/\!2,\uparrow}c^{\dagger}_{p_{\downarrow}\!+\!q\!/\!2,\downarrow}
 c_{p_{\downarrow}\!-\!q\!/\!2,\downarrow}c_{p_{\uparrow}\!+\!q\!/\!2,\uparrow}
 \punc{,}
\end{align}
where $\xi_{\vec{p}}=p^2/2-\mu$ is the dispersion, $\mu$ is the chemical
potential, and we adopt atomic units $\hbar=m=k_{\text{B}}=1$. A general
momentum dependent interaction $g(\vec{p},\vec{q})$ acts between the two
species, where $\vec{p}$ is the incoming relative momentum and $\vec{q}$
represents the momentum transfer. The Stoner Hamiltonian of electrons
interacting with a contact interaction would be recovered with
$g(\vec{p},\vec{q})=g$. To study the onset of ferromagnetic order, two
\emph{ab initio}, complementary, and mutually consistent tools have emerged:
the analytical order by disorder approach~\cite{Conduit08}, and Quantum
Monte Carlo~\cite{Conduit09,Pilati10}. We now extend both methods to probe
and understand the consequences of finite range interactions.

\subsection{Functional integration}

\begin{table}
 \begin{tabular}{l|c|c|c}
  &&Separable&Reciprocal\\
  &Solid state&cold atoms&cold atoms\\
  &$g(\vec{p},\!\vec{q})/g\!=$&$g(\vec{p},\!\vec{q})/g\!=$&$g(\vec{p},\!\vec{q})/g\!=$\\
  \hline
  Functional&
   $(1+b^{2}q^{2})^{-\!1}$&
   $1\!+\!2 ar_{\text{e}}(p^{2}\!+\!q^{2})$&
   $[1\!-\!2ar_{\!\text{e}}(p^{2}\!+\!q^{2})]^{\!-\!1}$\\
  Mean-field&
   $1$&
   $1+\frac{12}{5}k_{\text{F}}a k_{\text{F}}r_{\text{e}}$&
   Numerical\\
  Fluctuation&
   $(1\!+\!2k_{\text{F}}^{2}b^2)^{\!-\!1}$&
   $1+8 k_{\text{F}}a k_{\text{F}}r_{\text{e}}$&
   $(1\!-\!8k_{\text{F}}a k_{\text{F}}r_{\!\text{e}})^{\!-\!1}$
   \\\hline
  MF trans.&
   $g\nu_{\text{F}}=1$&
   $k_{\text{F}}a\!=\!\frac{\pi}{2}\!-\!\frac{3\pi^{\!2}\!k_{\text{F}}r_{\!\text{e}}}{5}$&
   Numerical
 \end{tabular}
 \caption{The rescaling of the interaction strength in the mean-field and
   fluctuation contributions to the free energy on introducing finite range
   interactions for the solid state and cold atoms cases. The bottom row shows
   the expected interaction strength of the magnetic transition in the
   mean-field approximation.}
 \label{tab:FormsOfPotential}
\end{table}

The functional integral formalism calculates the quantum partition function
expressed as a coherent state field integral
\begin{align}
&\mathcal{Z}=\!\int\mathcal{D}\psi\text{exp}\Bigg[-\!\!\!\!
\sum_{p,\sigma=\left\{\uparrow,\downarrow\right\}}\!\!\!\!
\overline{\psi}_{p,\sigma}(-\cmplxi\omega+\xi_{\vec{p}\sigma})\psi_{p,\sigma}\neweqnline
&-\!\!\!\!\sum_{p_{\uparrow},p_{\downarrow},q}
\!\!\!\!g(\vec{p}_{\uparrow}\!-\!\vec{p}_{\downarrow},\vec{q})
\overline{\psi}_{\!p_{\uparrow}\!-\!q\!/\!2,\uparrow}\overline{\psi}_{\!p_{\downarrow}\!+\!q\!/\!2,\downarrow}
\psi_{\!p_{\downarrow}\!-\!q\!/\!2,\downarrow}\psi_{\!p_{\uparrow}\!+\!q\!/\!2,\uparrow}\!\Bigg]\!\punc{,}
\label{eq:partition}
\end{align}
where the field $\psi$ describes a two component Fermi gas.  We use the
four-momentum notation $p=\{\omega,\vec{p}\}$ with Matsubara frequencies
$\omega$ and inverse temperature $\beta=1/T$.

We decouple the quartic interaction term with a Hubbard-Stratonovich
transformation into the full vector magnetization $\vecgrk{\phi}$ and
density channel $\rho$~\cite{Conduit08}. This decoupling scheme is sensitive
to the spin spiral instability driven by transverse magnetic fluctuations,
and simultaneously the p-wave superconducting instability driven by
longitudinal magnetic fluctuations.  The action is now quadratic in the
Fermionic degrees of freedom, and after integrating them out we recover the
quantum partition function
$\mathcal{Z}=\int\mathcal{D}\vecgrk{\phi}\mathcal{D}\rho\exp(-S)$ with an
action
\begin{align}
 S\!=\!\tr\left[ \vecgrk{\phi} \hat{g} \vecgrk{\phi} \!-\!\rho \hat{g}\rho \right]\!
 -\!\tr\ln\left[(\hat{\partial}_{\tau}\!+\!\hat{\xi}_{\alpha}\!
 +\!\hat{g}\rho)\mathbf{I}\!-\!\hat{g}\vecgrk{\phi}\cdot\vecgrk{\sigma}\right]\!\punc{,}
 \label{eq:phiaction}
\end{align}
where $\hat{g}$ is the operator form of the potential. We expand the action
to quadratic order in fluctuations of $\rho$ and $\vecgrk{\phi}$ around
their putative saddle-point values $\rho_{0}$ and
$\vec{M_{\vec{Q}}}=M(\cos\vec{Q}\!\cdot\!\vec{r},\sin\vec{Q}\!\cdot\!\vec{r},0)$,
where $\vec{Q}$ is the spiral wave vector.  A gauge transformation renders
the magnetization uniform and modifies the dispersion as
$\xi_{\vec{p}\sigma}=p^2/2+\sigma\sqrt{(\vec{p}\!\cdot\!\vec{Q})^2+(gM)^2}-\mu$.
After integrating over magnetization and density fluctuations the free
energy is
\begin{align}
&F=\!\sum_{\vec{p},\sigma}\!\frac{p^{2}}{2m_{\sigma}}
n(\xi_{\vec{p}\sigma})
\!+\!\!\!\sum_{\vec{p}_{\uparrow},\vec{p}_{\downarrow}}\!\!
g(\vec{p_{\uparrow}}\!-\!\vec{p_{\downarrow}},\vec{0})
n(\xi_{\vec{p_{\uparrow}}\uparrow})
n(\xi_{\vec{p_{\downarrow}}\downarrow})\neweqnline
&-\!\!\!\!\!
\sum_{
\substack{
\vec{p}_{1}+\vec{p}_{2} \\
=\vec{p}_{3}+\vec{p}_{4}
  }
}
\!\!\!\!g(\!\vec{p}_{1}\!\!-\!\!\vec{p}_{3},\vec{p}_{1}\!\!-\!\!\vec{p}_{4})^2
\frac
{n(\xi_{\vec{p}_{1}\!\uparrow}\!)n(\xi_{\vec{p}_{2}\!\downarrow}\!)
[n(\xi_{\vec{p}_{3}\!\uparrow}\!)\!+\!n(\xi_{\vec{p}_{ 4}\!\downarrow}\!)]}
{\xi_{\vec{p}_{1}\uparrow}+\xi_{\vec{p}_{2}\downarrow}
-\xi_{\vec{p}_{3}\uparrow}-\xi_{\vec{p}_{4}\downarrow}}
\!\punc{.}\nonumber
\end{align}
The first term is the kinetic energy, the second the mean-field contribution
of the interactions, and the third is the fluctuation correction. The
momentum summation in the mean-field term introduces a weighted average that
can be performed analytically for certain potentials at low temperature.
The momentum summation in the fluctuation term is dominated by the
contributions at $p^2+q^2=4 k^2_{\text{F}}$~\cite{Conduit09} and therefore
the interaction potential can be approximated by $g(\vec{p},\vec{q})\mapsto
g(\sqrt{2}k_{\text{F}},\sqrt{2}k_{\text{F}})$. These two observations allow
us to simply rescale the interaction strength that appears in the mean-field
and fluctuation correction terms and afterwards treat them as pure contact
interactions. The validity of the approximation will be verified in the cold
atom section.

We now enumerate the interaction potentials that we adopt for describing the
solid state and cold atom gas.
\begin{description}
\item[Solid state] We use the Coulomb interaction $g\e{-r/b}/4\pi b^2 r$ with
screening length $b$,
whose Fourier transform $g(\vec{p},\vec{q})=g/(1+b^2q^2)$ depends only on
the momentum transfer $\vec{q}$.
\item[Cold atoms] The T-matrix that describes the Feshbach resonance can be
  modeled by several potentials~\cite{Phillips98,Pethick02,Conduit09}. Here
  we concentrate on two: the separable form used by Pekker that facilitates
  the momentum summation~\cite{Pekker11,Kohstall12};
  $g(\vec{p},\vec{q})=(2k_{\text{F}}a/\pi\nu_{\text{F}})[1+2ar_{\text{e}}(p^{2}+q^{2})]$;
  and to establish a direct contact with the screened Coulomb interaction a
  reciprocal form (a Taylor expansion of the former)
  $g(\vec{p},\vec{q})=(2k_{\text{F}}a/\pi\nu_{\text{F}})/[1-2
  ar_{\text{e}}(p^{2}+q^{2})]^{-1}$.  Both potentials depend on the energy
  in the center-of-mass frame that appears in the scattering amplitude. Here
  the interaction strength is analogous to the scattering length
  $k_{\text{F}}a$, the effective range is $r_{\text{e}}$, and
  $\nu_{\text{F}}$ is the density of states at the Fermi surface.
\end{description}
In \tabref{tab:FormsOfPotential} we summarize how the interaction potentials
rescale the mean-field and fluctuation correction potentials after
performing the momentum summations.


\subsection{Quantum Monte Carlo}

The fluctuation corrections included in the analytical formalism represent a
subset of all possible contributions to the free energy. To gauge the
effectiveness of our analytical formalism we perform Diffusion Monte Carlo
calculations with the \textsc{casino} program~\cite{Needs10}. The approach
is a refinement of that used in previous studies of itinerant
ferromagnetism~\cite{Ceperley80,Conduit09,Pilati10}. This method optimizes a
trial wave function at zero temperature to yield the exact ground state
energy, subject only to a fixed node approximation, and thus neatly
complements the analytics. We use a variational wave function $\psi=\e{-J}D$
that is a product of a Slater determinant, $D$, that takes full account of
the Fermion statistics, and a Jastrow factor $J$ to include further
correlations.  We used a screened Coulomb repulsion $g\e{-r/b}/4\pi b^2r$
that exactly reflects the potential used in the analytics.

The Slater determinant consists of plane-wave spinor orbitals containing
both spin-up and spin-down electrons, $D=\det(\{\psi_{\vec{k}\in
  k_{\text{F},\uparrow}}, \bar{\psi}_{\vec{k}\in
  k_{\text{F},\downarrow}}\})$. Fixing the spin Fermi surfaces
$\{k_{\text{F}, \uparrow},k_{\text{F}, \downarrow} \}$ sets the
magnetization. For computational efficiency we factorize the Slater
determinant into an up and a down-spin determinant~\cite{Needs10}.  Provided
that the orbitals of the minority spin state are the lowest energy orbitals
of those in the majority spin state,~\cite{Roothaan60}, this gives the state
whose total spin is $S_{\text{tot}}^{\text{z}}$.

The Jastrow factor, $J$, accounts for electron-electron correlations. It
consists of the polynomial and plane-wave expansions in electron-electron
separation proposed in Ref.~\cite{Drummond04}. To further optimize the wave
function the orbitals in the Slater determinant were evaluated at
quasiparticle positions related to the electrons through a polynomial
backflow function~\cite{Rios06} that partially relieves the fixed node
approximation.  The trial wave functions were optimized using QMC methods
using VMC, backflow, and Diffusion Monte Carlo.  Twist averaging was
employed to remove finite size effects.

\section{Phase diagram}

\begin{figure}
 \includegraphics[width=1.\linewidth]{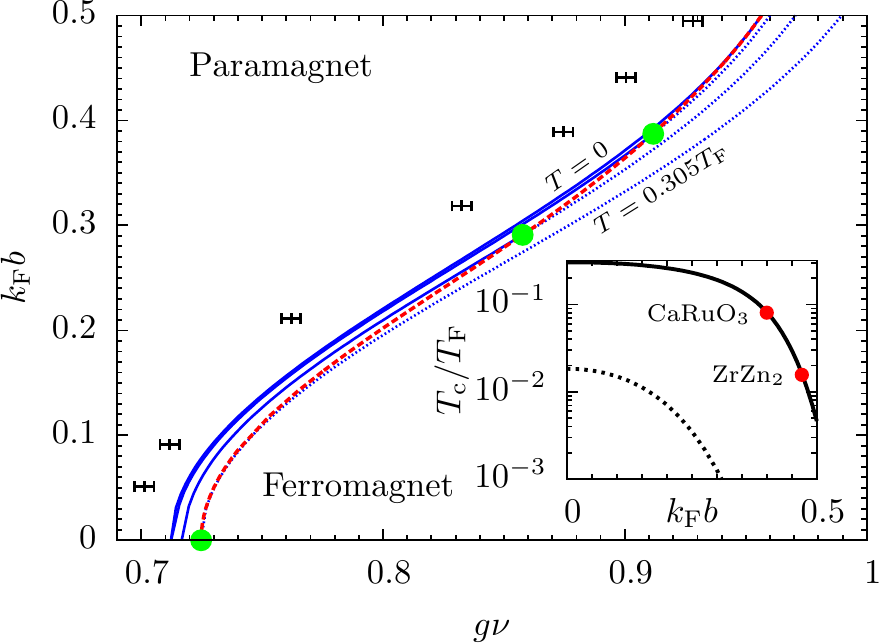}
 \caption{(Color online): Phase boundaries at temperatures $T \in \{
   0,0.1,0.2,0.305\} T_{\text{F}}$ with a screened Coulomb interaction of
   strength $g \nu$ and range $k_{\text{F}}b$. The ferromagnetic transition
   is first order (solid blue curves) at low $k_{\text{F}}b$, separated from
   the second order at larger $k_{\text{F}}b$ (dotted blue curves) by the
   green dots. The dashed red line is the locus of tricritical
   temperatures. The black points with error bars show the DMC $T=0$
   estimates for the phase boundary. The inset shows the tricritical
   temperature (solid black line) and the p-wave superconducting critical
   with $k_{\text{F}}b$, the red dots denote the CaRuO$_{3}$ family and
   ZrZn$_{2}$.}
 \label{fig:PhaseDiag}
\end{figure}

The analytical and computational tools developed calculate the free energy
of the electron gas with screened Coulomb interaction. We seek the
magnetization that minimizes the free energy to construct the phase
diagram. To orient the discussion we first start with the common contact
interaction at $k_{\text{F}}b=0$. In \figref{fig:PhaseDiag} fluctuation
corrections~\cite{Conduit08,Conduit09} drive the transition first order at
$g\nu\approx0.7$ (versus the mean-field second order transition at
$g\nu=1$). The importance of fluctuations is reduced as we increase the
temperature and ultimately the transition becomes second order at the
tricritical temperature $T_{\text{c}}=0.3 T_{\text{F}}$. This tricritical
point temperature is in agreement with previous studies of the contact
interaction (\cite{Conduit08,Conduit09}) but is an order of magnitude higher
than that seen in experiment.

At zero temperature, increasing the screening length diminishes the
interaction strength for the fluctuation correction as
$1/(1+4k^2_{\text{F}}b^2)$ (\tabref{tab:FormsOfPotential}). With the driving
force of the first order transition suppressed, the critical interaction
strength rises towards the mean-field critical interaction strength $g\nu
=1$. Nevertheless, the transition remains resolutely first order due to a
non-analytic contribution to the free energy of the form $M^4\log
\left(g^2M^2+T^2\right)$~\cite{Conduit09}. However, with increasing
$k_{\text{F}}b$ a slight temperature rise occludes the non-analyticity and
the transition reverts to second order, exemplified in the inset of
\figref{fig:PhaseDiag}, which shows how the tricritical temperature rapidly
reduces with rising screening length. Focusing on the CaRuO$_{3}$ family
with $k_{\text{F}}b=0.40$~\cite{Mazin97} and ZrZn$_{2}$ with
$k_{\text{F}}b=0.47$~\cite{Santil01}, the tricritical point temperature is
reduced to $0.08T_{\text{F}}$ and $0.02T_{\text{F}}$ respectively with
reasonable agreement to experiment~\cite{Hooper04,Borzi07,Wu11}.

We also study the magnetic transition at zero temperature using DMC. This
predicts a phase boundary with increasing screening length that is
quantitatively similar to the analytical predictions. Here the Monte Carlo
is in better agreement with the analytical prediction compared with previous
DMC studies \cite{Conduit09,Ceperley80,Pilati10} that employed a square
potential with uncontrolled effective screening length.

A Landau expansion with the zero range system demonstrates that the first
order ferromagnetic transition is always accompanied by a spin spiral
phase~\cite{Conduit09,Conduit13}. To incorporate the screening length we
take the same Landau expansion~\cite{Conduit13}, and rescale the interaction
parameters according to \tabref{tab:FormsOfPotential}. This reveals that the
spiral phase persists in a thin strip of order $\Delta g\nu \approx 4\times
10^{-4}$ between the first order phase boundary and the Lifshitz line,
terminating at the tricritical point. The transition temperature of an
instability to a p-wave superconductor was calculated by
Ref.~\cite{Conduit13} and the associated interaction strengths can be
rescaled following the prescription in \tabref{tab:FormsOfPotential}. In the
inset of \figref{fig:PhaseDiag} we find that the peak superconducting
transition drops with screening length even more rapidly than the
tricritical temperature, obscuring the instability in typical materials.

\subsection{Ultracold atomic gas}

\begin{figure}
 \includegraphics[width=1.\linewidth]{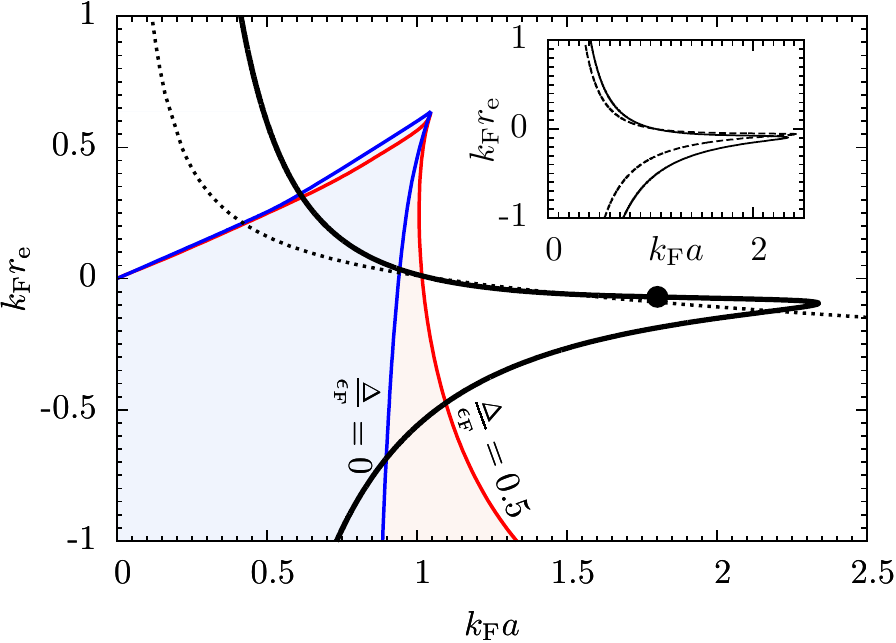}
 \caption{(Color online) The phase diagram for a $T=0$ cold atom gas with
   pseudopotential interaction strength $k_{\text{F}}a$ and effective range
   $k_{\text{F}}r_{\text{e}}$. The two shaded regions denote zero loss (shaded blue), and
   a loss rate less than $0.5\epsilon_{\text{F}}$ (shaded red). The thick black line
   represents the transition to the ferromagnetic state for the separable
   pseudopotential (whose fluctuation correction rescaling is zero at the
   black dot), and the dotted line the reciprocal pseudopotential
   transition. The inset compares the phase diagram for the full fluctuation
   corrections with separable pseudopotential (solid black line) to one obtained using the
   $g(\vec{p},\vec{q})\mapsto g(\sqrt{2}k_{\text{F}},\sqrt{2}k_{\text{F}})$
   approximation (dashed black line).}
 \label{fig:PhaseDiagCAG}
\end{figure}

As in the solid state, finite range interactions alter the nature of the
ferromagnetic transition. At zero effective range and zero temperature, the
separable and reciprocal forms for the pseudopotential are identical so in
\figref{fig:PhaseDiagCAG} both magnetic transitions concur with previous
studies that employed contact
interactions~\cite{Conduit08,Conduit10}. Sweeping the effective range
through zero raises the rescaled interaction strength, reducing the critical
interaction strength. At the negative effective range
$k_{\text{F}}r_{\text{e}}=-1/8k_{\text{F}}a$, the scaling of the fluctuation
term for the separable pseudopotential is zero, recovering the mean-field
free energy for which the transition occurs at $k_{\text{F}}a=5\pi/7$. At
more negative effective range the fluctuation term is restored, thus
reducing the critical interaction strength. In the case of the reciprocal
pseudopotential, at large effective range the rescaled interaction strengths
scale as $1/k_{\text{F}}r_{\text{e}}$, leading to a suppression of
interaction effects and a quenching of the ferromagnetic transition at
$k_{\text{F}}r_{\text{e}}\leq-0.68$. Although here the separable and
reciprocal pseudopotentials lead to different phase behavior, this regime is
not only beyond the effective range approximation where higher order terms
would be important, but furthermore in a regime where we will now determine
that losses dominate.

In a cold atom gas, to generate the repulsive interaction experimentalists
exploit a Feshbach resonance between the free atoms in the Fermi sea and the
bound state of two atoms to generate a positive scattering length.  However,
as the bound state is necessarily lower in energy there is an instability to
pair formation. The loss rate has been studied from both the
paramagnetic~\cite{Pekker11} and polaronic~\cite{Kohstall12} standpoints. To
determine whether the ferromagnetic or pairing instability dominates, we
follow Ref.~\cite{Pekker11} we start from \eqnref{eq:partition} and we
decouple this interaction using a Hubbard-Stratonovich transformation into
the Feshbach molecule channel $\Delta_{\vec{q}} \!= \!\sum_{\vec{k}} \!c_{
  \vec{k}\downarrow} c_{ \vec{q}-\vec{k}\uparrow }$. At leading order the
resulting Lagrangian is
\begin{align}
|\Delta_{\omega, \vec{q}}|^2
\!\underbrace{
\!\left(\!\frac{1}{g (\vec{0},\!\vec{q})}\!
+\!\!\!\int\!\!\! \frac{\text{d}^3 \vec{p}}{(\!2\pi\!)^3}\!
\frac{n_{\text{F}}(\xi_{\vec{p}+\! \frac{\vec{q}}{2}\uparrow})\!
+\!n_{\text{F}}(\xi_{\vec{p}-\!\frac{\vec{q}}{2}\downarrow})\!-\!1  }
{\text{i} \omega  - \xi_{\vec{p}+\frac{\vec{q}}{2} \uparrow} - \xi_{\vec{p}-\frac{\vec{q}}{2} \downarrow}}\!\right)
}_{C^{-1}(\omega, \vec{q})}\!\!\punc{.}
\end{align}
After regularizing the above integral~\cite{Pekker11}, and specializing to
an attractive potential with finite range $r_{\text{e}}$, we obtain an
expression for the pairing susceptibility $C(\omega, \vec{q})$
\begin{align}
&\left[ \frac{1}{g (\vec{0},\!\vec{q})} \!+ \!\frac{\text{i} m}{4\pi}\!
\sqrt{\!m\!\left(\!\text{i} \omega\!+\!2 \epsilon_{\text{F}} \!-\! \frac{
q^2}{4m }\!\right)}\!-\! \frac{m^2 r_{\text{e}}}{8\pi} \left(\! \text{i} \omega\!+\!2 \epsilon_{\text{F}} \!- \!\frac{q^2}{4 m }\!\right)  \!\nonumber\right. \\
&\left.+\! \int \!\!\frac{\text{d}^3 \vec{p}}{(2\pi)^3} \frac{n_{\text{F}}(\xi_{\vec{p}+\vec{q}/2 \uparrow}) \!+\! n_{\text{F}}(\xi_{\vec{p}-\vec{q}/2\downarrow}) }{\text{i} \omega  - \xi_{\vec{p}+\vec{q}/2,\uparrow} - \xi_{\vec{p}-\vec{q}/2,\downarrow}}\right]^{-1}\punc{.}
\end{align}
The imaginary component of the pairing susceptibility pole represents the
bound state pairing rate. We find generally that the maximal pairing rate
occurs for $\vec{q}=\vec{0}$.

In \figref{fig:PhaseDiagCAG} we overlay the magnetic transition with lines
of equal pairing rate $\Delta$. Pairing rate reduces with increasing
positive effective range as the molecules become more tightly bound. This
leads to a window in the phase diagram where the system is both magnetized
and there are no losses. However, at negative effective range the losses
occur on a time-scale $\sim 0.1 \,\text{ms}$ which is significantly shorter
than the trap crossing time $\sim 1\,\text{ms}$, so large magnetic domains
cannot be formed. To date all Fermionic mixtures used in cold atom gas
experiments have negative effective ranges so are not suitable for observing
magnetic correlations \cite{Grimm}, however a polar molecule gas with strong
dipolar interactions does display large positive effective range
\cite{Shi12} and a positive $s$-wave scattering length and so presents an
opportunity to observe ferromagnetic phenomena.

The inset of \figref{fig:PhaseDiagCAG} confirms that the phase boundary
obtained by rescaling the interaction strength of the fluctuation
contribution with the approximation $1+2ar_{\text{e}}(p^{2}+q^{2})\mapsto
1+8k_{\text{F}}ak_{\text{F}}r_{\text{e}}$ conforms with the phase boundary
resulting from the exact momentum summation. This verifies the rescaling
approximations given in \tabref{tab:FormsOfPotential}.

\section{Discussion}

The analytical fluctuation correction formalism with a contact repulsion has
successfully demonstrated how quantum fluctuations drive not only a first
order ferromagnetic transition, but also a spin spiral and a p-wave
superconducting instability. The finite ranged interactions reduce the
fluctuation corrections so dramatically suppress the tricritical point
temperature from $0.3 T_{\text{F}}$ to $0.02 T_{\text{F}}$. This is
quantitatively similar to the $T_{\text{c}}\approx0.02T_{\text{F}}$ seen in
ZrZn$_2$~\cite{Hooper04,Borzi07,Wu11}. Finite ranged interactions suppress
the transverse magnetic fluctuations that occluded quantum criticality. With the tricritical temperature now reduced this system allows for the observation of quantum critical phenomena, such as those envisaged by Hertz.

The idealized Stoner Hamiltonian can also be studied in a cold atom gas.
The positive effective finite range interaction, found in a polar molecular
gas, eliminates the pairing permitting the observation of ferromagnetic
order.

{\it Acknowledgments:} The authors are grateful to Stefan Baur, Andrew
Green, Jesper Levinsen, Pietro Massignan, and Stephen Rowley for fruitful
discussions. CVK acknowledges the financial support of the EPSRC, and GJC
from Gonville \& Caius College.

\end{document}